\documentclass[a4paper,12pt]{article}

\usepackage{amssymb,bm,mathrsfs,bbm,amscd}
\usepackage[titletoc]{appendix}
\usepackage{dsfont}
\usepackage{filecontents}
\usepackage[colorlinks=true, citecolor=blue, linkcolor=red, urlcolor=blue]{hyperref}
\usepackage[tbtags]{amsmath,nccmath}
\usepackage[mathscr]{euscript}
\usepackage{xcolor}

\makeatletter
\newcommand*{\Xbar}{}%
\DeclareRobustCommand*{\Xbar}{%
  \mathpalette\@Xbar{}%
}
\newcommand*{\@Xbar}[2]{%
  \sbox0{$#1\mathrm{X}\m@th$}%
  \sbox2{$#1X\m@th$}%
  \rlap{%
    \hbox to\wd2{%
      \hfill
      $\overline{%
        \vrule width 0pt height\ht0 %
        \kern\wd0 %
      }$%
    }%
  }%
  \copy2 %
}
\makeatother
\usepackage{lastpage,graphicx}
\usepackage [top=1.5in , bottom=1.5in, left=1.3in, right=1.2in] {geometry}

\usepackage{enumitem}
\usepackage{relsize}
\usepackage{tikz}
\usepackage{pgfplots}\pgfplotsset{compat=newest}
\usepackage{caption}
\usepackage[utf8]{inputenc}
\usepackage{lastpage}
\usepackage{titlesec}
   \titleformat*{\section}{\large\bfseries}
   \titleformat*{\subsection}{\bfseries}
\usepackage[english]{babel}
\usepackage{fancyhdr}
\usepackage{bigints}

\usepackage{natbib}
\usepackage{amsthm}

\theoremstyle{plain}
\newtheorem{thm}{Theorem}[section]
\newtheorem{cor}[thm]{Corollary}
\newtheorem{lem}[thm]{Lemma}
\newtheorem{prop}[thm]{Proposition}
\theoremstyle{definition}

\newtheorem{ass}{Assumption}[section]
\theoremstyle{remark}

\numberwithin{equation}{section}

\def\({\left(}
\def\){\right)}
\def\[{\left[}
\def\]{\right]}
\newtheoremstyle{theorem}{\topsep}{\topsep}%
     {}
     {}
     {}
     {}
     {.5em}
     {\thmname{{\bfseries #1}}\thmnumber{ #2}\thmnote{ #3}.}

\theoremstyle{theorem}

\newtheoremstyle{remark}{\topsep}{\topsep}%
     {}
     {}
     {\it}
     {}
     {.5em}
     {\thmname{{ #1}}\thmnumber{ #2}\thmnote{ #3}.}
\theoremstyle{remark}

\usepackage{amssymb}

\begin{document}

\thispagestyle{empty}
   ~~ \\[-1cm]
\begin{center}
                                \Large{\textbf{Malliavin differentiability of fractional Heston-type model and applications to option pricing}} \\[5mm]

         \normalsize{ \textbf{Marc Mukendi Mpanda}} \\[3mm]
         \normalsize{Department of Decision Sciences}\\
         \small{University of South Africa, P. O. Box 392, Pretoria, 0003. South Africa}\\
         \small{mpandmm@unisa.ac.za}\\

\end{center}

\vspace{0.1cm}
 \begin{abstract}

\noindent
This paper defines fractional Heston-type (\emph{fHt}) model as an arbitrage-free financial market model with the infinitesimal return volatility described by the square of a single stochastic equation with respect to fractional Brownian motion with Hurst parameter $H\in(0,1)$. We extend the idea of \citet{alos2008malliavin} [Alos, E., \& Ewald, C. O. (2008). Malliavin differentiability of the Heston volatility and applications to option pricing. Advances in Applied Probability, 40(1), 144-162.] to prove that \emph{fHt} model is Malliavin differentiable and deduce an expression of expected payoff function having discontinuity of any kind.  Some simulations of stock price process and option prices are performed.\\

\noindent
\textbf{Keywords:} Fractional Heston-type model, Fractional Brownian motion, Fractional Cox-Ingersoll-Ross process and Malliavin differentiability.
\end{abstract}

\section{Introduction}
Allowing volatility to be stochastic in a financial market model was one of the great achievement in the history of quantitative finance. This yields stochastic volatility modelling that was previously discussed by \citet{heston1993closed} and several other researchers to overcome shortfalls in the standard Black-Schole model (See e.g. \citet{alos2019volatility} for a summary). In the sense of \citet{heston1993closed}, the stock price process is described by a geometric Brownian motion
$$
dS_t = \eta S_t dt + \sqrt{Y_t} S_t dB_t,
$$

\noindent
where $\eta$ and $\sqrt{Y_t}$ represent the drift and stochastic variance of the infinitesimal return $X_t:= \log S_t$. The stochastic process $(Y_t)_{t\geq 0}$ takes the form of standard Cox-Ingersoll-Ross process that satisfies the following stochastic differential equation:
$$
dY_t = \theta(\mu -  Y_t)dt + \nu\sqrt{Y_t} d\tilde{B}(t).
$$

\noindent
The parameter  $\theta$  represents the speed of reversion of the stochastic process $(Y_t)_{t\geq 0}$ towards its long-run mean $\mu$ and the parameter $\nu$ represents the volatility of $(Y_t)_{t\geq 0}$. The Brownian motions $(B_t)_{t\geq 0}$ and $(\tilde{B}_t)_{t\geq 0}$ are assumed to be correlated. This model is well known in the literature as the \emph{``Heston model''}.\\

\noindent
The standard Heston model comes with three main drawbacks: (1) the spot volatility is driven by a standard Brownian motion which does not display memory. New findings show roughness in volatility time-series (see e.g. \citet{comte1998long}, \citet{chronopoulou2010hurst} for long-range dependency or \citet{gatheral2018volatility}, \citet{livieri2018rough} and subsequent results for short range dependency known as ``rough volatility'').  (2) Perfect calibration may not be possible as the stochastic volatility parameters are constants. It was proven that dependent parameters reduce the calibration error sensibly (See e.g. \citet{benhamou2010time}) and (3) the analytical solution of option price is very complex especially for exotic payoff functions.\\



\noindent
This paper addresses these issues by defining the stock price process as a geometric Brownian motion $(S_t)_{t\geq 0}$ that satisfies the following stochastic differential equation: \\[-10mm]

\begin{equation}\label{Eq1-1}
dS_t = \eta S_t dt + \sigma(Y_t) S_t dB_t,
\end{equation}

\noindent
where $\sigma(Y_t)$ represents the volatility of the infinitesimal log-return $dX_t:=dS_t/S_t$ with $(Y_t)_{t\geq 0}$ a fractional Cox-Ingersoll-Ross (\textit{fCIR}) process that captures both long and short range dependency. We opt for the definition of \citet{mishura2018fractional} and describe the stochastic process $(Y_t)_{t\geq 0}$ as

\begin{equation}\label{Eq1-2}
Y_t(\omega) = Z^2_t(\omega)\mathbf{1}_{[0,\tau(\omega))}, ~~~\forall t \geq 0, ~~\omega\in\Omega,
\end{equation}
\noindent
where the stochastic process  $(Z_t)_{t\geq 0}$ is referred to a general form of \textit{fCIR} process that satisfies the following differential equation:
\begin{equation}\label{Eq1-3}
dZ_t = \frac{1}{2}\Big(f(t,Z_t)Z_t^{-1}dt + \nu dW_t^H\Big), ~~\nu>0,
\end{equation}

\noindent
and $\tau$ is the first time the process $(Z_t)_{t\geq 0}$ hits zero defined by
\begin{equation}\label{Eq1-4}
\tau(\omega) = \inf\big\{t>0 : Z_t(\omega) = 0\big\}.
\end{equation}

\noindent
In (\ref{Eq1-3}), the function $f(t,z)$ represents the drift of the volatility process $(Y_t)_{t\geq 0}$ and the stochastic process $(W^H_t)_{t\geq 0, \, H\in(0,1)}$ is well-known as fractional Brownian motion (\textit{fBm}) of Hurst parameter $H$ defined as a centered Gaussian process with covariance function

\begin{equation}\label{Eq1-5}
\mathds{E}\big[W^H_t W^H_s\big]=\frac{1}{2}\Big(t^{2H}+s^{2H}-\mid t-s\mid^{2H}\Big), ~~\forall s,t \geq 0.
\end{equation}

\noindent
Recall that \emph{fBm} can be represented in terms of stochastic integral in at least three different ways: time representation, Harmonisable representation and Volterra representation (See \citet{nourdin2012selected} for more details). In what follows, we shall consider the Volterra representation of \emph{fBm} given by
\begin{equation}\label{Eq1-6}
W^H_t = \int_{0}^{t}\kappa_H(s,t)dV_t,
\end{equation}
where $(V_t)_{t\in [0,T]}$ is a standard Brownian motion and where $\kappa_H(s,t)$ is a square integrable  kernel defined by

\begin{equation}\label{Eq1-7}
\kappa_H(t,s) = \frac{(t-s)^{H-\frac{1}{2}}}{\Gamma(H+\frac{1}{2})} \,\,{}_2 \mathbf{F}_1\Big(H-\frac{1}{2};\frac{1}{2}-H;H+\frac{1}{2};1-\frac{t}{s}\Big) \mathbf{1}_{[0,t]}(s), ~\forall s\in [0,t],
\end{equation}

\noindent
with $\Gamma(\cdot)$ and ${}_2 \mathbf{F}_1(a,b,c;d)$ the gamma and Gaussian hypergeometric functions respectively. The standard Brownian motions $(B_t)_{t\in [0,T]}$ and $(V_t)_{t\in [0,T]}$  are assumed to be correlated, that is, there exists $\rho \in [-1,1]$ such that $ \mathds{E}\big[B_tV_t\big] = \rho t.$ This means that there exists a Brownian motion $(\tilde{V}_t)_{t\in [0,T]}$ independent to $(V_t)_{t\in [0,T]}$, that is $\mathds{E}\big[V_t,\tilde{V}_t\big] = 0$, such that
\begin{equation}\label{Eq1-8}
B_t = \rho V_t + \sqrt{1-\rho^2} \tilde{V}_t.
\end{equation}

\noindent
Now taking into consideration the risk-free asset process $(A_t)_{t\geq 0}$, the fractional Heston-type (\textit{fHt}) model is given by the following system

\begin{equation}\label{Eq1-9}
\begin{cases}
dA_t = rA_tdt, \\[1mm]
dX_t = \eta dt + \sigma(Y_t) dB_t, \\[1mm]
Y_t = Z_t^2 \mathbf{1}_{[0,\tau(\omega)]}\\[3mm]
dZ_t = \frac{1}{2}f(t,Z_t)Z_t^{-1}dt + \frac{1}{2}\nu dW_t^H\\[1mm]
W^H_t = \mathlarger\int_{0}^{t}\kappa_H(s,t)dV_t \\[1mm]
B_t = \rho V_t + \sqrt{1-\rho^2} \tilde{V}_t,
\end{cases}
\end{equation}

\noindent
The existence of stochastic process $(Z_t)_{t\geq 0}$ in (\ref{Eq1-3}) was previously discussed by \citet{nualart2002regularization}. They proposed that for $H<1/2$, the drift function $g(t,z) := f(t,z)z^{-1}$ must satisfy the linear growth condition and for $H>1/2$, $g(t,z)$ must verify the H\"{o}lder continuity condition. \\

\noindent
\noindent
Particular cases of \textit{fHt} model (\ref{Eq1-9}) has been previously investigate by \citet{alos2017fractional}, \cite{bezborodov2019option} and \citet{mishura2019option} for $H >1/2$.\\

\noindent
One can use the same idea of \citet[Theorem 4]{bezborodov2019option} to show that the \textit{fHt} model is free of arbitrage. In this paper, we also show that both stock price and fractional volatility processes are Malliavin differentialble through their approximating sequences, and deduce the expected payoff function.\\

\noindent
The remainder of this paper is structured as follows: Section 2 constructs an approximating sequences of stock prices and \textit{fCIR} processes. The Malliavin differentiability within the \textit{fHt} model is discussed in Section 3. Finally, Section 4 derives the expected payoff function and perform some simulations of option prices.

\section{Approximating sequences in \textit{fHt} model.}

The main purpose of introducing approximating sequences of both fractional volatility and stock price processes relies on the their positiveness. The following theorems discuss the positiveness of $(Z_t)_{t\geq 0}$ and before this, we consider the following assumption.

\begin{ass}\label{Ass2-1}~~\\[-6mm]
	\begin{itemize}
		\item[\textbf{(i)}] The function $g: [0, \infty) \times (0, \infty) \to (-\infty,  \infty)$ defined by
		$g(t,z):=f(t,z)/z$ is continuous and admits a continuous partial derivative with respect to $x$ on $(0, \infty)$.
		\item[\textbf{(ii)}] for any $T >0$, there exists  $z_{T} >0$  such that
		$$f(t, z) > 0 \mbox{  for all }  0 < t \leq T \mbox{ and } 0 \leq z \leq z_{T}.$$
	\end{itemize}
\end{ass}
\noindent
Under this assumption, the following theorems were proved by \citet{mishura2018fractional} or \citet{mpanda2020generalisation}.

\begin{thm}\label{Thm2-1} Let $(Z_t)_{t\geq 0}$ be a stochastic process that verifies \textit{(\ref{Eq1-3})} with $H >\frac{1}{2}$ and $f:[0, \infty) \times [0, \infty)$ is a continuous function that satisfies Assumption \ref{Ass2-1}. Then
	 $$\mathds{P}(\tau = \infty) = 1,$$
where  $\tau(\omega) = \inf\{t>0 : Z_t(\omega) = 0\}.$
\end{thm}

\begin{thm}\label{Thm2-2} Consider for each $k>0$, the stochastic process $(Z_t^{(k)})_{t\geq 0}$ defined by \\[-6mm]
\begin{eqnarray*}
	Z_t^{(k)}= \left\{ \begin{array}{ll}
		Z_0 + \mathlarger\int_{0}^{t} \dfrac{f_k(t,Z_s^{(k)})}{Z_s^{(k)}}ds+\dfrac{\nu}{2} W^H_t & \mbox{ if }  t<\tau^{(k)}(\omega)\\
		0 & \mbox{ otherwise,}
	\end{array}
	\right.
\end{eqnarray*}
where $\tau^{(k)}(\omega) = \inf\{t\geq 0 : Z_t^{(k)}(\omega) = 0\}.$ Then for any $T>0$ and $H<1/2$,			        		
	$$\mathds{P}(\omega\in\Omega:\tau^{(k)}(\omega) > T) \to 1 ~\mbox{ as } ~ k \to \infty.$$
\end{thm}

\subsection{Approximating sequences of $(Z_t)_{t\geq 0}$}

Inspired by \citet{alos2008malliavin}, we construct an approximating sequence $(Z^\epsilon_t)_{t\geq 0,\,\epsilon>0}$ of the \textit{fCIR} process that satisfies the following differential equation:

\begin{equation}\label{Eq2-1}
dZ^\epsilon_t = \frac{1}{2}f(t,Z^\epsilon_t)\Lambda_\epsilon(Z^\epsilon_t)dt + \frac{\sigma}{2}dW_t^H, ~~~~ Z_0^\epsilon = Z_0 > 0,
\end{equation}

\noindent
where the function  $\Lambda_\epsilon(z)$ in (\ref{Eq2-1}) is defined by
\begin{equation}\label{Eq2-2}
\Lambda_\epsilon(z) = (z \mathbf{1}_{\{ z >0 \}} + \epsilon)^{-1}.
\end{equation}
\noindent
It is easy to verify that $\Lambda_\epsilon(z) > 0$ for all $\epsilon > 0$. As a straight consequence, the drift of $(Z^\epsilon_t)_{t\geq 0,\,\epsilon>0}$ is also positive. In addition,  $\lim_{z\to 0}\Lambda_\epsilon(z)  = \epsilon^{-1}$, $\lim_{z\to \infty}\Lambda_\epsilon(z)  = 0$ and

\begin{equation}\label{Eq2-3}
\Lambda'_\epsilon(z) =
\left\{
\begin{array}{rl}
0, & \hbox{if} ~~  z < 0\\
-\frac{1}{(z+\epsilon)^2}, & \hbox{if} ~~ z \geq 0
\end{array}
\right.
\end{equation}

\noindent
The next step is to show that for every $t\geq 0$, the sequence $Z^{\epsilon}_t$ converges to $Z_t$ in $L^p$ as $\epsilon\to 0$.

\begin{prop}\label{Prop2-3} The sequence of estimated random variables $Z^{\epsilon}_t$ converges to $Z_t$ in $L^p(\Omega)$ for all $p\geq 1$.
\end{prop}

\noindent
\textbf{Proof.}	 \\[-7mm]
\begin{itemize}
	\item[] \textbf{Case 1.}  $H = 1/2$. This case was discussed previously by \citet[Proposition 2.1]{alos2008malliavin} and can be easily extended to the case where $\Lambda_\epsilon(z)$ is defined by (\ref{Eq2-2}).
	\item[] \textbf{Case 2.} For $H > 1/2$, the dominated convergence theorem shall be applied.  Firstly, we need to show the pointwise convergence of the approximated stochastic process  $(Z^\epsilon_t)_{t\geq 0}$ towards $(Z_t)_{t\geq 0}$, that is  $\lim_{\epsilon\to 0} Z^\epsilon_t = Z_t$. For this, let $\tau_\epsilon(\omega) = \inf \{t\geq 0 : Z_t(\omega) \leq \epsilon\}$ be the first time the process $(Z_t)_{t\geq 0}$ hits $\epsilon$ . Since the sample paths of the stochastic process $(Z_t)_{t\geq 0}$ are positive everywhere almost surely as in Theorem \ref{Thm2-1}, then $\mathds{P}(\omega\in\Omega:\tau_0 = \infty) = 1$ as and consequently, $\lim_{\epsilon\to 0} \tau_\epsilon = \infty$ almost surely. \\[-4mm]
	
	\noindent
	Next, denote $(Z_t^{\tau_\epsilon})_{t\in [0,\tau_\epsilon]}$ the stochastic process $(Z_t)_{t\geq 0}$ up to stopping time $\tau_\epsilon$. Then, for all $t\in [0,\tau_\epsilon]$ and using the definition of  $\Lambda_\epsilon(z)$ given by (\ref{Eq2-3}), $Z_t^{\tau_\epsilon} = Z^\epsilon_t$ almost surely when $\epsilon\to 0$ since the drift function $f(t,z)$ is monotonic. \\[-4mm]

	\noindent
	Again, the positiveness of $(Z_t)_{t\geq 0}$ means that $\lim_{\epsilon\to 0} Z_t^{\tau_\epsilon} = Z_t ~a.s.$ We may conclude that $\lim_{\epsilon\to 0} Z_t^{\tau_\epsilon} =\lim_{\epsilon\to 0} Z_t^{\epsilon}  = Z_t$ almost surely and for all $t\geq 0$.  \\
	
	\noindent
	On the other hand, the result from \citet[Theorem 3.1]{hu2008singular} shows that for a fixed $T>0$ and for all $p\geq 1$, \\[-4mm]
	$$
	\mathds{E}\big[\sup_{t\in[0,T]} \big|Z_t\big|^p\big] = C < \infty,
	$$
	\noindent
	where $C = C(p,H, \gamma, \beta, T, Z_0)$ is a non-random constant taking the form \\[-10mm]
	
	$$
	C = C_1(1+Z_0) \exp \Bigg[C_2 \Big(1+\big|\big| W^H \big|\big|^{\frac{\gamma}{\beta(\gamma-1)}}\Big)\Bigg],
	$$
	\noindent
	where $\beta \in (\frac{1}{2}, H)$, $\gamma>\frac{2\beta}{2\beta - 1}$,  $C_1 = C_1(\gamma,\beta,T)$ and $C_2 = C_2(\gamma,\beta,T)$ are nonrandom constants depending on parameters $\gamma,\beta,T$,  and \\[-3mm] $$||W^H|| = \sup_{s \geq 0, \, t \leq T}\Biggl\{\frac{|W^H_s - W^H_t|}{|s-t|^\beta}\Biggr\}.$$
	\noindent
	This result also implies that
	$$
	\mathds{E}\big[\sup_{t\in[0,T]} \big|Z_t^\epsilon\big|^p\big] = C(p,H, \gamma, \beta, T, Z_0) < \infty.
	$$
	\noindent
	It follows that $\sup_{t\in [0,T]}\big\{\big|Z_t^\epsilon(\omega)\big|\} \in L^p(\Omega)$ which yields the desired $L^p$ convergence.
	
	
	\item[] \textbf{Case 3.} For $H < 1/2$, we consider a sequence of an increasing drift functions $f_k(t,z), ~k\in \mathbb{N}$ and define the stochastic process $(Z_t^{(\epsilon,k)})_{t\geq 0}$ as follows: \\[-3mm]
	
	\begin{eqnarray*}
		Z_t^{(\epsilon,k)}= \left\{ \begin{array}{ll}
			Z_0 + \dfrac{1}{2}\mathlarger\int_{0}^{t}  f_k\(t,Z_s^{(\epsilon,k)}\) \Lambda\(Z_s^{(\epsilon,k)}\)ds+\dfrac{\nu}{2} W^H_t & \mbox{ if }  t<\tau^{(k)}(\omega)\\
			0 & \mbox{ otherwise,}
		\end{array}
		\right.
	\end{eqnarray*}
	
	\noindent
	where $\Lambda(z)$ is defined by (\ref{Eq2-2}) and $\tau^{(k)}(\omega) = \inf\{t\geq 0 : Z_t^{(\epsilon,k)}(\omega) = 0\}$ is the first time that the stochastic process $(Z_t^{(\epsilon,k)})_{t\geq 0}$ hits zero. If we now define $\tau^{(\epsilon,k)}(\omega) = \inf\{t\geq 0 : Z_t^{(\epsilon,k)}(\omega) \leq \epsilon\}$ be the first time the process $(Z_t^{(\epsilon,k)})_{t\geq 0}$ hits $\epsilon$, then from Theorem \ref{Thm2-2}, for any fixed $T>0$, $\mathds{P}(\omega\in\Omega:\tau^{(\epsilon,k)} > T) \to 1$ as $k\to \infty$. This implies that $\lim_{(\epsilon,k)\to (0,\infty)}  \tau^{(\epsilon,k)} = \tilde{T} > T$ a.s. This is because the process $(Z_t^{(\epsilon,k)})_{t\geq 0}$ remains positive up to time $\tilde{T}$ which is not necessary equal to infinity unlike the previous case. \\[-4mm]
	
	\noindent
	After using similar arguments of \textbf{Case 2}, one may conclude that  $\lim_{\epsilon\to 0} Z_t^{\tau_\epsilon} =\lim_{\epsilon\to 0} Z_t^{\epsilon}  = Z_t$ for all $t\in [0, \tilde{T}]$. Next, we need to show that $ \mathds{E}\big[\sup_{t\in[0,T]} \big|Z_t\big|^p\big] < \infty$. To achieve this, we borrow some ideas from \citet{mishura2019fractional}. \\[-4mm]

	\noindent
	Firstly, let $\tilde{Z}_0$ be a small positive value less than the initial value $Z_0$ such that $0<\tilde{Z}_0<Z_0$ and let $\tau_1 = \tau_1(\epsilon,\omega)$ be the last time the stochastic process $(Z_t^\epsilon)_{t\geq 0, \,\epsilon>0}$ hits (or before hits) $\tilde{Z}_0$, that is, \\[-6mm]
	
	\begin{equation}\label{Eq2-4}
	\tau_1(\epsilon,\omega) = \sup\{t\geq 0: Z_t^\epsilon(\omega) \geq \tilde{Z}_0, ~ \forall t\in [0,T]\}.
	\end{equation}
	\noindent
	Technically, there exists a constant $M\geq 2$ such that $\tilde{Z}_0 = \frac{Z_0}{M}$. Now we can consider two cases: $t\in[0, \tau_1]$ and $t\in(\tau_1, T]$.\\
	
	\noindent
	\textbf{Case 3.1:} $t\in[0, \tau_1]$. By triangle  inequality, we have \\[-6mm]
	
	\begin{equation}\label{Eq2-5}
	\begin{aligned}
	|Z^\epsilon_t|^p =& \Big|Z_0 + \frac{1}{2}\int_{0}^{t}f(s,Z^\epsilon_s)\Lambda_\epsilon(Z^\epsilon_s)ds + \frac{\nu}{2}W_t^H\Big|^p\\
	\leq &  \Bigg(Z_0 + \frac{1}{2}\Bigg|\int_{0}^{t}f(s,Z^\epsilon_s)\Lambda_\epsilon(Z^\epsilon_s)ds\Bigg| + \frac{\nu}{2}\big|W_t^H\big|\Bigg)^p\\
	\leq &\Bigg(Z_0 + \frac{1}{2}\int_{0}^{t}\Big|f(s,Z^\epsilon_s)\Lambda_\epsilon(Z^\epsilon_s)\Big|ds + \frac{\nu}{2}\big|W_t^H\big|\Bigg)^p.
	\end{aligned}
	\end{equation}
	
	\noindent
	By applying the Callebaut's inequality theorem, it will be easy to show that for all $p\geq 1$,\\[-4mm]
	\begin{equation}\label{Eq2-6}
	\begin{aligned}
	&\Bigg(Z_0 + \frac{1}{2}\int_{0}^{t}\Big|f(s,Z^\epsilon_s)\Lambda_\epsilon(Z^\epsilon_s)\Big|ds + \frac{\nu}{2}\big|W_t^H\big|\Bigg)^p\\
	& ~~~~~~~~~~ \leq 3^{p}\Bigg(Z_0^p + \(\frac{1}{2}\int_{0}^{t}\Big|f(s,Z^\epsilon_s)\Lambda_\epsilon(Z^\epsilon_s)\Big|ds\)^p + \(\frac{\nu}{2}\big|W_t^H\big|\)^p\Bigg).\\
	\end{aligned}
	\end{equation}
	From (\ref{Eq2-4}), we may deduce that $Z_t^\epsilon \geq  \tilde{Z}_0 > 0$, with $t$ on $[0, \tau_1]$. This yields $\Lambda_\epsilon (Z_t^\epsilon) < MZ_0^{-1}, ~M\geq 2$ and
	
	\begin{equation}\label{Eq2-7}
	\begin{aligned}
	\int_{0}^{t}\Big|f(s,Z^\epsilon_s)\Lambda_\epsilon(Z^\epsilon_t)\Big| ds  \leq  & \(\frac{M}{Z_0}\) \int_{0}^{t}\Big|f(s,Z^\epsilon_s)\Big| ds.
	\end{aligned}
	\end{equation}
	
	\noindent
	Since the drift function satisfies the linear growth condition, this means there exists a positive constant $k$ such that $f(t,z) \leq k(1+|z|)$. It follows that
	\begin{equation}\label{Eq2-8}
	\begin{aligned}
	\int_{0}^{t}\Big|f(s,Z^\epsilon_s)\Big| ds \leq \int_{0}^{t}\Big|k(1 + |Z^\epsilon_s|)\Big| ds \leq k \(T + \int_{0}^{t} |Z^\epsilon_s| ds\).
	\end{aligned}
	\end{equation}
	
	\noindent
	Inequalities (\ref{Eq2-6}), (\ref{Eq2-7}) and (\ref{Eq2-8}) yield the following:\\
	$$
	|Z^\epsilon_t|^p \leq 3^{p}\Bigg(Z_0^p + \(\frac{kM}{2 Z_0}\)^p\(T + \int_{0}^{t} |Z^\epsilon_s| ds\)^p + \(\frac{\nu}{2}\)^p\big|W_t^H\big|^p\Bigg).
	$$

	\noindent
	On the other hand, recall that $|W_t^H| < \sup_{s\in[0,T]} |W_s^H|<\infty$ (See e.g. \citet{nourdin2012selected}) and since
	$$
	\(T + \int_{0}^{t} |Z^\epsilon_s| ds\)^p \leq 2^p\(T^p + \int_{0}^{t} |Z^\epsilon_s|^p ds\),
	$$
	
	\noindent
	then it follows that
	\begin{equation*}
	\begin{aligned}
	|Z^\epsilon_t|^p &\leq (3Z_0)^p + \(\frac{3kMT}{Z_0}\)^p + \(3\nu\)^p\sup_{s\in[0,T]} \big|W_s^H\big|^p +\(\frac{3kM}{Z_0}\int_{0}^{t} |Z^\epsilon_s| ds\)^p.\\
	&\leq  (3Z_0)^p + \(\frac{3kT}{Z_0}\)^p + \(4\nu\)^p\sup_{s\in[0,T]} \big|W_s^H\big|^p +\(\frac{3k}{Z_0}\int_{0}^{t} |Z^\epsilon_s| ds\)^p.\\
	\end{aligned}
	\end{equation*}
	
	\noindent
	From the Gr\"{o}nwall-Bellman inequality theorem, we obtain\\
	\begin{equation*}
	\begin{aligned}
	|Z^\epsilon_t|^p &\leq \bigg( (3Z_0)^p + \(\frac{3kMT}{Z_0}\)^p + \(4\nu\)^p\sup_{s\in[0,T]} \big|W_s^H\big|^p \bigg)\exp\(\(\frac{3kM}{Z_0}\)^pt\)\\
	& \leq \bigg( (3Z_0)^p + \(\frac{3kMT}{Z_0}\)^p\bigg)\exp\(\(\frac{3kM}{Z_0}\)^pT\) \\
	& ~~~~~~~~~~ + \bigg( \(4\nu\)^p\sup_{s\in[0,T]} \big|W_s^H\big|^p \bigg)\exp\(\(\frac{3kM}{Z_0}\)^pT\)\\
	\end{aligned}
	\end{equation*}
	
	\noindent
	which can be shortly written as $|Z^\epsilon_t|^p \leq C,$ where $C = C(r, k, T, Z_0, \nu, H)$ is a non-random constant in parameters $r, k, T, Z_0, \nu$ and $H$ taking the following form
	$$
	C \leq C_1 + C_2 \sup_{s\in[0,T]} \big|W_s^H\big|^p,
	$$
	
	\noindent
	with $C_1 = C_1(p, k,T,Z_0)$ and $C_2 = C_2(p, k,T,Z_0,\nu)$ are non-random constants defined  respectively by
	\begin{equation}\label{Eq2-9}
	C_1 = (3Z_0)^p\bigg( 1 + \(\frac{kMT}{Z_0^2}\)^p\bigg)\exp\(\(\frac{3kM}{Z_0}\)^pT\)
	\end{equation}
	
	\noindent
	and
	\begin{equation}\label{Eq2-10}
	C_2 = (4\nu)^{p} \exp\(\(\frac{3kM}{Z_0}\)^pT\).
	\end{equation}
	
	\noindent
	\textbf{Case 3.2:} $t\in(\tau_1, T],$ with $T>\tau_1> 0.$ Define
	$$ \tau_2 = \tau_2(\epsilon,\omega) = \sup\{s \in (\tau_1,t): |Z^\epsilon_s(\omega)|<\tilde{Z}_0\}.
	$$
	\noindent
	Then we have:
	\begin{equation}\label{Eq2-11}
	\begin{aligned}
	|Z^\epsilon_t|^p &\leq |Z^\epsilon_t - Z^\epsilon_{\tau_2}|^p + |Z^\epsilon_{\tau_2}|^p \\[1mm]
	& \leq Z_0^p + |Z^\epsilon_t - Z^\epsilon_{\tau_2}|^p \\
	& \leq Z_0^p +  \(\frac{1}{2}\)^p\bigg|\int_{\tau_2}^{t}f(s,Z^\epsilon_s)\Lambda_\epsilon(Z^\epsilon_t)ds + \nu\big(W_t^H - W^H_{\tau_2}\big)\bigg|^p\\[2mm]
	& \leq Z_0^p +  \(\int_{\tau_2}^{t} \Big|f(s,Z^\epsilon_s)\Lambda_\epsilon(Z^\epsilon_t)\Big|ds\)^{p} + (2\nu)^p\Big(\big|W_t^H\big|^p + \big|W^H_{\tau_2}\big|^p\Big).\\
	\end{aligned}
	\end{equation}
	
	\noindent
	As previously, the integral in the last inequality of (\ref{Eq2-11}) can be expressed as follows
	$$
	\int_{0}^{t}\Big|f(s,Z^\epsilon_s) \Lambda_\epsilon(Z^\epsilon_t) \Big|ds \leq \frac{k}{Z_0} \(T + \int_{0}^{t} |Z^\epsilon_s| ds\), ~~~\forall t \in [0,T].
	$$
	\noindent
	On the other hand, we may observe that
	$$
	\big|W_t^H\big|^p + \big|W^H_{\tau_2}\big|^p \leq 2 \sup_{s\in [0,T]} |W^H_s|^p.
	$$
	\noindent
	It follows that,\\[-2mm]
	\begin{equation*}
	\begin{aligned}
	|Z^\epsilon_t|^p &\leq Z_0^p + \(\frac{2kT}{Z_0}\)^p  + \(\frac{2k}{Z_0}\int_{0}^{t} |Z^\epsilon_s|^r ds\)^p + 2(2\nu)^p\sup_{s\in [0,T]} |W^H_s|^p\\
	&\leq (3Z_0)^p + \(\frac{3kMT}{Z_0}\)^p + \(4\nu\)^p\sup_{s\in[0,T]} \big|W_s^H\big|^p +\(\frac{3kM}{Z_0}\int_{0}^{t} |Z^\epsilon_s| ds\)^p .\\
	\end{aligned}
	\end{equation*}
	\noindent
	From this expression, we may also conclude that $|Z^\epsilon_t|^p \leq C,$ where $C = C(C_1,C_2)$ where $C_1$ and $C_2$ are a non-random constants defined by (\ref{Eq2-9})  and (\ref{Eq2-10}) respectively. This shows that $ \mathds{E}\big[\big|Z_t^\epsilon\big|^p\big] < \infty$ and consequently, $ \mathds{E}\big[\sup_{t\in[0,T]} \big|Z_t\big|^p\big] < \infty$. This concludes the proof of the proposition.         \hfill{$\Box$}\\
\end{itemize}

\begin{cor}\label{Cor4-5} Fix $p\geq 1$. If $\sigma(y)$ satisfies the linear growth condition, then
	$$
	\lim_{\epsilon\to 0} \mathds{E} \[\sup_{t\geq 0}\big|\sigma(Y_t^\epsilon) - \sigma(Y_t)\big|^p\] = 0 ~~~a.s.
	$$	
\end{cor}

\noindent
\textbf{Proof.} This follows immediately from the previous proposition. \\

\noindent
\textbf{Remark.} One may use similar arguments of \citet{mishura2019fractional} to show that the stochastic process $(Z^\epsilon_t)_{t\geq 0,\, \epsilon >0}$ is strictly positive almost surely for all $H\in(0,1)$. Consequently, it is also well suitable for rough volatility processes, that is, fractional volatility process with $H<1/2$.

\subsection{Approximating sequences of stock price process}

With $(Z^\epsilon_t)_{t\geq 0, \, \epsilon>0}$, let us construct the approximating sequence $(S_t^\epsilon)_{t\geq 0, \,\epsilon>0}$ of the stock price process $(S_t)_{t\geq 0}$ defined by the following geometric Brownian motion:
\begin{equation}\label{Eq2-12}
dS_t^\epsilon = \eta S_t^\epsilon dt + \sigma(Y_t^\epsilon) S_t^\epsilon dB_t,
\end{equation}
\noindent
where
$$
Y_t^\epsilon = (Z^\epsilon_t)^2,
$$
\noindent
with $(Z_t^\epsilon)_{t\geq 0, \,\epsilon>0}$ the approximating sequence that satisfies (\ref{Eq2-1}). The solution to (\ref{Eq2-12}) is unique and can be found by using the standard It\^o formula and it is given by. Next step is to show that $S_t^\epsilon$ converges to $S_t$ in $L^p, \, p\geq 1$.

\begin{prop}\label{Prop4-7} Set $X_t := \log S_t$ and $X^{\epsilon}_t := \log S^{\epsilon}_t$. Then the sequence $X^{\epsilon}_t$ converges to $X_t$ in $L^p(\Omega)$ for all $p\geq 1$.
\end{prop}

\noindent
\textbf{Proof.} Firstly, we have from It\^o formula that\\[-3mm]
\begin{equation}\label{Eq2-13}
X_t^\epsilon = X_0 + \eta t - \frac{1}{2}\int_{0}^{t}\sigma^2(Y^\epsilon_s)ds + \int_{0}^{t}\sigma(Y^\epsilon_s)dB_s,
\end{equation}

\noindent
where $X_0 := \log S_0$. Then for some non-random constant $C>0$, one may have: \\[-2mm]
\begin{equation*}
\begin{split}
\mathds{E}\[\sup_{t\geq 0} \big| X^\epsilon_t - X_t \big|^p\] \leq & \frac{C}{2^p} \mathds{E} \[ \sup_{t\geq 0}\Bigg|\int_{0}^{t}\(\sigma^2(Y^\epsilon_s) - \sigma^2(Y_s)\)ds \Bigg|^p\] \\[2mm]
& ~ + C \mathds{E}\[ \sup_{t\geq 0} \Bigg|\int_{0}^{t}\(\sigma(Y^\epsilon_s) - \sigma(Y_s)\) dB_s\Bigg|^p\] \\
\end{split}
\end{equation*}

\noindent
Set
$$
\mathbb{T}_1 := \mathds{E} \[ \sup_{t\geq 0}\Bigg|\int_{0}^{t}\(\sigma^2(Y^\epsilon_s) - \sigma^2(Y_s)\)ds \Bigg|^p\]
$$
\noindent
and
$$
\mathbb{T}_2 := \mathds{E}\[ \sup_{t\geq 0} \Bigg|\int_{0}^{t}\(\sigma(Y^\epsilon_s) - \sigma(Y_s)\) dB_s\Bigg|^p\].
$$

\noindent
Then it follows firstly that $\mathbb{T}_1 \to 0$ from Corollary \ref{Cor4-5}. To analyse convergence of  $\mathbb{T}_2$, the Burkholder-Davis-Gundy inequality can be used and one may deduce that
$$
\mathbb{T}_2 \leq c(p) \mathds{E}\[ \sup_{t\geq 0} \Bigg|\int_{0}^{t}\(\sigma(Y^\epsilon_s) - \sigma(Y_s)\) ds\Bigg|^\frac{p}{2}\],
$$

\noindent
which also converges to zero from Corollary \ref{Cor4-5}. It follows that
$$\lim_{\epsilon\to 0} \sup_{t\geq 0} \big|	X^\epsilon_t - X_t \big|^p = 0, ~~\forall p>0$$

\noindent
that implies the desired $L^p$ convergence of $X^\epsilon_t$  to $X_t$ and $S^\epsilon_t$  to $S_t$. \\

\noindent
\textbf{Remarks.}
\begin{itemize}
	\item [(1)] The approximated stochastic volatility and stock price processes will be compulsory for $H \leq 1/2$ and optional for $H>1/2$. However, for the sake of consistency, we shall use the approximated sequences (\ref{Eq2-1}) with $\epsilon = 0$ for $H>1/2$  and with $\epsilon > 0$ for $H \leq 1/2$.
	\item [(2)] For the simulations of stock price process, one may use the Euler-Maruyama approximation scheme. This can be done by considering the time interval $[0,T]$ that is subdivided into $N$ sub-intervals of equal length such that $0=t_0, t_1,\cdots,t_N=T$ with $t_i = iT/N$ and the lag $\Delta t = T/N$. The estimated stock price at time $t_i$ denoted by $(\hat{S}_{t_i})_{i = 1,\cdots,N}$ and the volatility $(\hat{Y}_{t_i})_{i = 1,\cdots,N}$ are respectively given by \\[-4mm]
	
	\begin{equation}\label{Eq2-14}
	\begin{cases}
	\hat{S}_{t_{i+1}} = \hat{S}_{t_i} \Bigg(1 + \eta \Delta t + \sigma(\hat{Y}_{t_i}) \, \(\rho \Delta V_{t_i} + \sqrt{1 - \rho^2} \Delta \tilde{V}_{t_i}\)\Bigg)  \\[3mm]
	\hat{Y}_{t_i} = \hat{Z}_{t_i}^2 1_{[0,\tau(\omega)]}\\[3mm]
	\hat{Z}_{t_{i+1}} = \hat{Z}_{t_{i}} +  \dfrac{1}{2}\mathlarger{\int}_{0}^{t_{i+1}}f(s,\hat{Z}_{s})\Lambda(\hat{Z}_{s})ds + \frac{1}{2}\nu \Delta W^H_{t_{i+1}}.
	\end{cases}
	\end{equation}
	where $\Delta V_{t_{i}} = V_{t_{i+1}} - V_{t_{i}}, ~~ \Delta \tilde{V}_{t_{i}} = \tilde{V}_{t_{i+1}} - \tilde{V}_{t_{i}}$ and $\Delta W^H_{t_{i+1}} = W^H_{t_{i+1}} - W^H_{t_{i}}$ are respectively the increment of Brownian motions $V_{t\in[0,10]}$, $\tilde{V}_{t\in[0,T]}$ and \textit{fBm} $W^H_{t\in[0,T]}$. \\
\end{itemize}	
	
\noindent
As an illustrative example, the following figures represent 10 sample paths of the stock price process on the interval $[0,T]$ with $N=1000, ~\rho = 0.6, ~X_0 = 100, ~\eta = r = 0.05, ~\nu = 0.1, ~\sigma(\hat{Y}_{t_i}) = 0.8 \hat{Y}_{t_i} + 0.1$. The drift of the fractional volatility process is defined by
	\begin{equation}\label{Eq2-15}
	f(t,y) =  \frac{\sigma^2}{2} \Big(1 - e^{-2\kappa t}\Big) + \kappa(c - y^2), \,\,\, t\geq 0, y \geq 0,
	\end{equation}
	\noindent
	with $\kappa = 1, ~ c=2$. For $H> 1/2$, we use $\epsilon = 0$ (See e.g. Figures 2.3 and 2.4) and for $H\leq 1/2$, we set $\epsilon = 0.01$ as shown in Figures 2.1 and 2.2.
	
	\begin{center}
		\begin{tabular}{cc}
			\includegraphics[width=2.4in, height=2.2in]{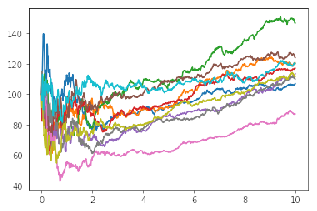} & \includegraphics[width=2.5in, height=2.2in]{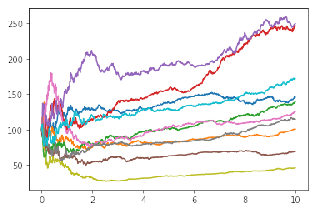} \\[-3mm]
			Figure 2.1: $H=0.15, ~\epsilon = 0.01$ & Figure 2.2: $H=0.5, ~\epsilon = 0.01$
		\end{tabular}
	\end{center}
	
	\begin{center}
		\begin{tabular}{cc}
			\includegraphics[width=2.4in, height=2.2in]{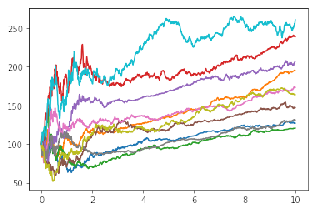} & \includegraphics[width=2.5in, height=2.2in]{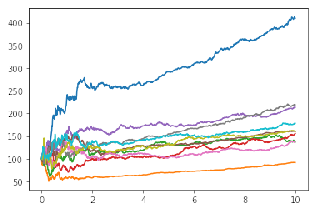}  \\[-3mm]
			Figure 2.3: $H=0.65, ~\epsilon = 0$ & Figure 2.4: $H=0.9, ~\epsilon = 0$
		\end{tabular}
	\end{center}

\section{Malliavin  differentiability}

\noindent
In what follows, we show  that the stochastic  processes $(Z_t)_{t\geq 0}$ and $(S_t)_{t\geq 0}$ are  Malliavin  differentiable with respect to the Brownian motions $(V)_{t\geq 0}$, $(\tilde{V})_{t\geq 0}$ and \textit{fBm} $(W^H_t)_{t\geq 0}$. We refer the reader to \citet{nualart2006malliavin} for a background in Malliavin calculus.

\subsection{Differentiability of the stochastic process $(Z_t)_{t\geq 0}$}

\begin{prop}\label{Prop3-1}
	
	Let $(Z^\epsilon_t)_{t\geq 0, \,\epsilon>0}$ be a stochastic process that verifies the stochastic differential equation (\ref{Eq2-1}) driven by a fBm $(W^H_t)_{t\in [0,T]}$ that takes the Volterra representation form given by
	$$
	W^H_t = \int_{0}^{t}\kappa_H(s,t) dB_s,
	$$
	
	\noindent
	where $(B_t)_{t\geq 0}$ is a standard Brownian motion and $\kappa_H(s,t)$ is a square integrable kernel given by (\ref{Eq1-7}). Assume that the drift function $f(t,z)$ is differentiable and define
	$$
	F_\epsilon(t,z) = \frac{\partial f(t,z)}{\partial z} \Lambda_\epsilon(z) + f(t,z)\Lambda'_\epsilon(z),
	$$
	\noindent
	where $\Lambda'_\epsilon(z)$ is defined by (\ref{Eq2-3}). Moreover, let $\EuScript{D}_u^B$ and  $\EuScript{D}_u^{W}$ be the Malliavin derivatives at the time $u\in [0,T]$ with respect to  $(B_t)_{t\geq 0}$  and  $(W^H_t)_{{t\geq 0}}$ respectively. Then it follows that  $Z^\epsilon_t \in \mathbb{D}^{1,p}$, \\[-1mm]
	
	\begin{equation}\label{Eq3-1}
	\EuScript{D}_u^B Z^\epsilon_t =  \frac{\nu}{2}\Bigg(\kappa_H(t,u) + \int_{u}^{t}\kappa_H(s,u) F_\epsilon(s,Z_s^\epsilon) \exp\Big(\int_{s}^{t}F_\epsilon(u,Z_u^\epsilon) du\Big)ds\Bigg) \mathbf{1}_{[0,t]}(u)
	\end{equation}
	\noindent
	and \\[-10mm]
	
	\begin{equation}\label{Eq3-2}
	\EuScript{D}_u^W Z^\epsilon_t =  \frac{\nu}{2}\Bigg(\exp\Big(\int_{s}^{t}F_\epsilon(u,Z_u^\epsilon) du\Big)\Bigg) \mathbf{1}_{[0,t]}(u).
	\end{equation}
	
\end{prop}

\noindent
\textbf{Proof.}  The Malliavin derivative $\EuScript{D}^B_u Z_t$ can be found as follows:\\[-4mm]	
\begin{equation*}
\begin{split}
\EuScript{D}^B_u Z_t^\epsilon = & \frac{1}{2}\int_{0}^{t} \EuScript{D}^B_u \(f(s,Z_s^\epsilon)\Lambda_\epsilon(Z_s^\epsilon)\) ds + \frac{\nu}{2}\EuScript{D}^B_u W^H_t\\[3mm]
= & \frac{1}{2}\int_{0}^{t} F_\epsilon(s,Z_s^\epsilon) \EuScript{D}^B_u Z_s^\epsilon ds + \frac{\nu}{2}\kappa_H(t,u)\mathbf{1}_{[0,t]}(u)\\[3mm]
\end{split}
\end{equation*}
The function $F_\epsilon(s,z)$ exists indeed since $\Lambda'_\epsilon(z)$ is well-defined for all Hurst parameter $H\in(0,1)$. Next, by letting $D_t = \EuScript{D}^B_u Z_t^\epsilon $, we obtain a Volterra integral equation given by \\[-8mm]

$$
D_t = \frac{1}{2}\int_{0}^{t} F_\epsilon(s,Z_s^\epsilon) D_s ds + \frac{\nu}{2}\kappa_H(t,u)\mathbf{1}_{[0,t]}(u),
$$
to which a solution is given by
$$
D_t = \frac{\nu}{2}\Bigg(\kappa_H(t,u) + \int_{u}^{t}\kappa_H(s,u) F_\epsilon(s,Z_s^\epsilon) \exp\Big(\int_{s}^{t}F_u du\Big)ds\Bigg) \mathbf{1}_{[0,t]}(u).
$$
\noindent
Since $D_t \in L^p(\Omega)$, then it follows that the stochastic process $Z^\epsilon_t \in \mathbb{D}^{1,p}$ from \citet{nualart2006malliavin}. The proof of (\ref{Eq3-2}) can be deduced in a similar way or by following the idea of \citet[Theorem 3.3]{hu2008singular}.  \hfill{$\Box$}\\[-3mm]

\newpage
\noindent
\textbf{Remarks}

\begin{enumerate}
\item [(1)] This proposition holds for all $H\in(0,1)$. However, for $H>1/2$ one may use directly $(Z_t)_{t\geq 0}$ given by (\ref{Eq1-3}) without going through its approximating sequence $(Z^\epsilon_t)_{t\geq 0, \, \epsilon>0}$ since the sample paths of $(Z_t)_{t\geq 0}$ are strictly positive everywhere almost surely as in Theorem \ref{Thm2-1}.
\item [(2)] As a straight consequence of Proposition \ref{Prop2-3}, we have
	$$
	\lim_{\epsilon\to 0}F_{\epsilon}(t,z) = F(t,z)
	$$
	where
	$$
	F(t,z) = \Big(\frac{\partial f(t,z)}{\partial z} - f(t,z)\Big) z^{-2}.
	$$
It follows from Proposition \ref{Prop3-1} that $Z_t \in \mathbb{D}^{1,p}$, and\\[-6mm]
	
	\begin{equation}\label{Eq4-22}
	\EuScript{D}_u^{\tilde{V}} Z_t = 0,
	\end{equation}
	\vspace{-8mm}
	\begin{equation}\label{Eq4-23}
	\EuScript{D}_u^V Z_t =  \frac{\nu}{2}\Bigg(\kappa_H(t,u) + \int_{u}^{t}\kappa_H(s,u) F(s,Z_s) \exp\Big(\int_{s}^{t}F(u,Z_u) du\Big)ds\Bigg) \mathbf{1}_{[0,t]}(u)
	\end{equation}
	\noindent
	and \\[-12mm]
	
	\begin{equation}\label{Eq4-24}
	\EuScript{D}_u^W Z_t =  \frac{\nu}{2}\Bigg(\exp\Big(\int_{s}^{t}F(u,Z_u) du\Big)\Bigg) \mathbf{1}_{[0,t]}(u).
	\end{equation}
\end{enumerate}

\subsection{Differentiability of the stock price process $(S_t)_{t\geq 0}$}

The following proposition shows that the stock price process and its estimation are Malliavin differentiable.

\begin{prop}\label{Prop4-8} Assume that the volatility $\sigma(y)$ is Lipschitz and differentiable. Then $S_t^\epsilon, X_t^\epsilon \in \mathbb{D}^{1,p}$  and for all $u\leq t$, we have
	\begin{equation}\label{Eq3-6}
	\EuScript{D}_u^{B} S_t^\epsilon = S_t^\epsilon\EuScript{D}_u^{B} X_t^\epsilon, ~~ \EuScript{D}_u^{V} S_t^\epsilon = S_t^\epsilon\EuScript{D}_u^{B} X_t^\epsilon ~\text{ and } ~ \EuScript{D}_u^{\tilde{V}} S_t^\epsilon = S_t^\epsilon\sqrt{1-\rho^2}\,\sigma(Y_t^\epsilon) \mathbf{1}_{[0,t]}(u),
	\end{equation}
	\noindent
	where \\[-10mm] 
	
	\begin{equation}\label{Eq3-7}
	\EuScript{D}_u^{B} X_t^\epsilon = \Bigg(\int_{u}^{t}\sigma'(Y_s^\epsilon) \EuScript{D}_u^{B} Y_s^\epsilon dB_s - \int_{u}^{t}\sigma(Y_s^\epsilon)\sigma'(Y_s^\epsilon) \EuScript{D}_u^{B} Y_s^\epsilon ds\Bigg)\mathbf{1}_{[0,t]}(u)
	\end{equation}

\noindent
and	\\[-4mm]

\begin{equation}\label{Eq3-8}
\begin{aligned}
\EuScript{D}_u^{V} X_t^\epsilon =& \Bigg(\rho \int_{u}^{t}\sigma'(Y_s^\epsilon) \EuScript{D}_u^V Y_s^\epsilon dV_s + \sqrt{1-\rho^2}\, \int_{u}^{t}\sigma'(Y_s^\epsilon) \EuScript{D}_u^V Y_s^\epsilon d\tilde{V}_s \\
								 & \qquad   - \int_{u}^{t}\sigma(Y_s)\sigma'(Y_s^\epsilon) \EuScript{D}_u^{V} \sigma(Y_s^\epsilon) ds\Bigg)\mathbf{1}_{[0,t]}(u)
\end{aligned}
\end{equation}



\noindent
In addition,

$$
\sup_{u,\,t \geq 0} \Big | \EuScript{D}_u X_t^\epsilon - \EuScript{D}_u X_t \Big| \to 0,
$$	

\noindent
where $\EuScript{D}_u$ represents a Malliavin derivative with respect to $B_t$, $V_t$ or $\tilde{V}$.

\end{prop}

\noindent
\textbf{Proof.} The equations (\ref{Eq3-6}) follows immediately from chain rule formula for Malliavin derivatives. Expressions of derivatives $\EuScript{D}_u^{B} X_t^\epsilon$ and $\EuScript{D}_u^{V} X_t^\epsilon$ are straight consequences of \citet[Theorem 1.2.4]{nualart2006malliavin}.

\begin{cor}\label{Cor4-9} The laws of both stock price process $(S_t)_{t\geq 0}$ and its log-return $(X_t)_{t\geq 0}$ are absolutely continuous.	
\end{cor}

\noindent
\textbf{Proof}. One may verify that $\big|\big|\EuScript{D}_u^{B} X_t\big|\big|_{L^2(\Omega)} > 0$ and $\big|\big|\EuScript{D}_u^{B} S_t\big|\big|_{L^2(\Omega)} > 0$ almost surely, then the absolutely continuity with respect to the Lebesgue measure on $\mathbb{R}$ follows immediately from \citet[Theorem 2.1.3]{nualart2006malliavin}. \\

\noindent
\textbf{Remark.} The Malliavin differentiability property of both stochastic volatility and stock price processes will be crucial for the derivation of the expected payoff function that will be discussed in the next chapter.

\section{Application to option pricing}


\noindent
The aim of this section is to derive the expected payoff function  $\mathbb{E}\[h(S_T)\]$ by using some results from Malliavin calculus and deduce its option price. We follow \citet{altmayer2015multilevel} closely. 

\subsection{The Expected Payoff function}


Let $h: \mathds{R} \to \mathds{R}$ be the payoff function that satisfies the following assumption.

\begin{ass}\label{Ass5-1}
	The payoff function $h: \mathds{R} \to \mathds{R}$ and its antiderivative denoted by $L(x)$ (such that $L(x) = h(x)$) are bounded and verify the Lipschitz condition.
\end{ass}

\begin{prop}\label{Prop5-1} $L(S_T) \in \mathbb{D}^{1,2}$.
\end{prop}

\noindent
\textbf{Proof.} Firstly, it is straightforwards to check that $\mathbb{E}[L^2(S_T)] < \infty$ since $L(x)$ also verifies the linear growth condition and the law of stock price process $(S_t)_{t\in [0,T]}$ are bounded almost surely. On the other hand, since $L$ verifies Assumption \ref{Ass5-1} and the sample paths of the stock price process $(S_t)_{t\in [0,T]}$ is absolutely continuous with respect to the Lebesgue measure on $\mathbb{R}$ (See Corollary \ref{Cor4-9}), then from the chain rule formula for Malliavin derivatives, we may deduce \\[-4mm]

		$$\EuScript{D}^V L(S_T) = L'(S_T) \EuScript{D}^V S_T = h(S_T) \EuScript{D}^V S_T.$$

\noindent
It follows that

\begin{equation*}
\begin{aligned}
\mathbb{E}\[\int_0^T\(\EuScript{D}^V_s L(S_T)\)^2ds\] &= \mathbb{E}\[\int_{0}^{T}\Big(h(S_T)\EuScript{D}^V_s S_T\Big)^2ds\]\\
& = \mathbb{E}\[h^2(S_T)\int_{0}^{T}\(\EuScript{D}^V_s S_T\)^2ds\] \\
& \leq \( \mathbb{E}\[h^4(S_T)\] \int_{0}^{T}\mathbb{E}\[\(\EuScript{D}^V_s S_T\)^4\]ds\)^{\frac{1}{2}} < \infty.\\
\end{aligned}
\end{equation*}

\noindent
The first inequality is due to Holder inequality and the finiteness of the last expression makes sense since $S_t \in \mathbb{D}^{1,2}$ as discussed previously. It follows that $||L||_{1,2} < \infty$ which concludes the proof. \hfill{$\Box$} \\

\noindent
As now $L(S_T)$ is Malliavin differentiable, then the following lemma that discusses the expected payoff follows.  					

\begin{lem}\label{PRP6-2} Let $h(x), ~ x\in \mathbb{R}$ be a  payoff function that satisfies Assumption \ref{Ass5-1} and denote $h(e^x) := g(x)$ with its antiderivative $G(x)$ that also satisfies the Lipschitz condition. Set \\[-4mm]
	\begin{equation}\label{Eq5-2}
	I_T := \frac{1}{T \sqrt{1-\rho^2}} \int_0^T \frac{1}{\sigma(Y_u)}d\tilde{V}.
	\end{equation}
	
	\noindent	
	Then \\[-12mm]
	
	\begin{equation}\label{Eq5-3}
	\mathbb{E}\[g(X_T)\] =  \mathbb{E} \Big[G(X_T) I_T\Big],
	\end{equation}
	\noindent
	and
	\begin{equation}\label{Eq5-4}
	\mathbb{E}\[h(S_T)\] =  \mathbb{E} \Bigg[\frac{L(S_T)}{S_T} \Big(1+ I_T\Big)\Bigg].
	\end{equation}
	\noindent
	where $X_T:=\log S_T$ and
	\begin{equation}\label{Eq5-5}
	L(S_T) = \int_{0}^{S_T}h(x)dx.
	\end{equation}
\end{lem}

\noindent
\textbf{Proof.}  We follow the idea of \citet{altmayer2015multilevel}. To establish the equality (\ref{Eq5-3}), we rewrite $\mathbb{E}[g(X_T)]$ as
$$
\mathbb{E}[g(X_T)] = \mathbb{E}\[\frac{1}{T}\int_0^Tg(X_T)du\] = \mathbb{E}\[\frac{1}{T}\int_0^T g(X_T)\EuScript{D}^{\tilde{V}}_u X_T \frac{1}{\EuScript{D}^{\tilde{V}}_u X_T}du\].
$$

\noindent
From Proposition \ref{Prop5-1}, we may deduce that $G(X_T) \in \mathbb{D}^{1,2}$ and
$$\EuScript{D}^{\tilde{V}} G(X_T) = g(X_T) \EuScript{D}^{\tilde{V}} X_T.$$

\noindent
We now obtain
$$
\mathbb{E}[g(X_T)] = \mathbb{E}\[\frac{1}{T}\int_0^T \EuScript{D}^{\tilde{V}} G(X_T) \frac{1}{\EuScript{D}^{\tilde{V}}_u X_T}du\].
$$

\noindent
In addition, from Proposition \ref{Prop4-8},
$$\EuScript{D}^{\tilde{V}}_u X_T = \sqrt{1-\rho^2}\sigma(Y_u) \mathbf{1}_{[0,t]}(u)$$
\noindent
and since the integral $\int_0^T \frac{1}{\sigma(Y_u)}du$ is well defined from Assumption \ref{Ass3-1}, then we have:

$$
\mathbb{E}[g(X_T)] = \mathbb{E}\[ \frac{G(X_T)}{T \sqrt{1-\rho^2}} \int_0^T \frac{1}{\sigma(Y_u)}d\tilde{V}_u\],
$$

\noindent
and defining $I_T$ by (\ref{Eq5-2}), we obtain (\ref{Eq5-3}). To establish (\ref{Eq5-4}), we rewrite the function $G(x)$ (which is the antiderivative of $g(x)$) as follows\\
$$
G(x) = \int_{0}^{x} g(u)du + C,
$$
where $C$ is a constant taking the form $C = \int_{0}^{1} h(u)du$ and by using the standard integration by part formula, one may obtain
$$
G(x) = \frac{L(e^x)}{e^x} + \int_{0}^{x} \frac{L(e^u)}{e^u} du.
$$

\noindent
With this setting, we have

\begin{equation*}
\begin{aligned}
\mathbb{E}[h(S_T)] & = \mathbb{E}[g(X_T)] \\
& = \mathbb{E} \Big[G(X_T) I_T\Big] \\
& =  \mathbb{E} \Bigg[\(\frac{L(S_T)}{S_T} + \int_{0}^{X_T} \frac{L(e^u)}{e^u} du\) I_T\Bigg] \\
& = \mathbb{E} \Bigg[\frac{L(S_T)}{S_T}I_T\Bigg] + \mathbb{E}\Bigg[\(\int_{0}^{X_T} \frac{L(e^u)}{e^u} du\) I_T\Bigg] \\
& =  \mathbb{E} \Bigg[\frac{L(S_T)}{S_T}I_T\Bigg] +  \mathbb{E}\Bigg[\frac{L(S_T)}{S_T}\Bigg] \\
& =  \mathbb{E} \Bigg[\frac{L(S_T)}{S_T} \Big(1 + I_T\Big)\Bigg]. \\
\end{aligned}
\end{equation*}



\noindent
We may use again the Euler-Maruyama approximation scheme to compute the expected payoff numerically. We may use the following approximations: \\[-4mm]

\begin{equation}\label{Eq6-3}
\begin{cases}
\hat{S}_{t_{i+1}} = \hat{S}_{t_i} \Bigg(1 + \eta \Delta t + \sigma(\hat{Y}_{t_i}) \, \(\rho \Delta V_{t_i} + \sqrt{1 - \rho^2} \Delta \tilde{V}_{t_i}\)\Bigg)  \\[3mm]
\hat{Y}_{t_i} = \hat{Z}_{t_i}^2 1_{[0,\tau(\omega)]}\\[3mm]
\hat{Z}_{t_{i+1}} = \hat{Z}_{t_{i}} +  \dfrac{1}{2}\mathlarger{\int}_{0}^{t_{i+1}}f(s,\hat{Z}_{s}) \Lambda (\hat{Z}_{s})ds + \frac{1}{2}\nu \Delta W^H_{t_{i+1}}\\[3mm]
\hat{I}_T =\dfrac{1}{T \sqrt{1-\rho^2}} \mathlarger{\sum}_{i = 0}^N \dfrac{1}{\sigma(\hat{Y}_i)}\Delta\tilde{V}_{t_i},
\end{cases}
\end{equation}

\noindent
with $0=t_0, t_1,\cdots,t_N=T$ with $t_i = iT/N$ and the lag $\Delta t = T/N$.

%
	


\subsection{Some simulations}


\noindent
\textbf{Pricing options with volatility taking the form of Ornstein-Uhlenbeck and standard \textit{fCIR} process}\\

\noindent
Firstly, we consider the stochastic process $(Z_t)_{t\geq 0}$ defined as a Ornstein-Uhlenbeck process, that is with $f(t,z) = -\theta z^2,$ where $\theta$ is a positive parameter, $\nu =2$ and $H>1/2$. Under these settings, one may recover the model discussed by \citet{bezborodov2019option} with $Y_t = Z_t^2$ instead. In this case, the volatility process will not be necessary positive almost surely since it violates the Assumption \ref{Ass2-1} and consequently the Theorems \ref{Thm2-1} and \ref{Thm2-2} do not apply. To compensate this, the volatility function $\sigma(y)$ is chosen to be strictly positive. \\

\noindent
In addition, we define the payoff function $h(x)$ as a combination of European and binary options with the same strike price $K$ and time to maturity $T$, that is  $h(S_T) = (S_T-K)_+ + \mathbf{1}_{S_T>K}$. It is easy to check that the strike price $K$ is a removable discontinuity of the payoff function $h$. In addition, the expression of $L(S_T)$ can be deduced from (\ref{Eq5-5}) as \\[-7mm]

\begin{eqnarray}
L(S_T)= \left\{ \begin{array}{ll}
\frac{1}{2} \Big[\big(S_T-K\big)\big(S_T-K+2\big)\Big] & \mbox{ if }  S_T\geq S\\
0 & \mbox{ otherwise.}
\end{array}
\right.
\end{eqnarray}


\noindent
We use the same parameters ($\eta=r=0.2, \,\theta = 0.6, \, T=1, H=0.6 $) with different forms of volatility process $\sigma(Y_t)$ of the infinitesimal return process $dS_t/S_t$ as in \citet{bezborodov2019option}. Since the \textit{fCIR} process of the form \ref{Eq1-2} and \ref{Eq1-3} cannot be used, we consider the direct form of the stochastic volatility $(Y_t)_{t\geq 0}$ driven by a \textit{fBm} represented by the Volterra stochastic integral (\ref{Eq1-5}) which can be discretised as follows: \\[-8mm]

\begin{equation}\label{Eq5-10}
W^H_{t_j} = \sum_{i = 0}^{j-1} \(\int_{t_{i-1}}^{t_i}\kappa_H(t_j,s)ds\)\delta V_i,
\end{equation}

\noindent
for all $j = 1, \cdots, N; ~ i=0, \cdots, j$ and where $\delta V_i = V_i - V_{i-1}$ is the increment of standard Brownian motion with $W^H_{t_0} = 0$. Here $\kappa_H(t_j,s)$ is a discretised square integrable kernel (\ref{Eq1-7}) given by \\[-4mm]

\begin{equation}\label{Eq5-11}
\kappa_H(t_j,s) = \frac{(t_j-s)^{H-\frac{1}{2}}}{\Gamma(H+\frac{1}{2})} \,\,{}_2 \mathbf{F}_1\Big(H-\frac{1}{2};\frac{1}{2}-H;H+\frac{1}{2};1-\frac{t_j}{s}\Big) \mathbf{1}_{[0,t_j]}(s), ~\forall s\in [0,t_j].
\end{equation}

\noindent
In this case, we observe that the values of option prices are not remarkably different for $\rho = 0$ and $H\geq 1/2$. The option prices are increasing or decreasing when $\rho$ is positive or negative respectively. \\


\noindent
Now, consider the fractional volatility process described by a standard \textit{fCIR} process, that is, with $f(t,z) = \mu - \theta z^2$ and correlation $\rho$ between infinitesimal returns and volatility, the option prices are simulated with $\rho = 0.5$ and $\mu = 0.1$.\\

\noindent
We do 100 trials for 500 simulations and 500 time-steps on the time interval $[0,1]$. We get the mean of option prices (that is, expected payoff function discounted by the net present value) with their corresponding coefficient of variations. Table \ref{Tab5-1} corresponds to the formula (\ref{Eq5-4}) and Table \ref{Tab5-2} to direct estimation of expected payoff function. \\

		\begin{table}[]
			\centering
			\caption{Option prices using Direct Estimations}
			\label{Tab5-1}
\vspace{-3mm}
			\resizebox{\columnwidth}{!}{%
				\begin{tabular}{|c|c|c||c|c||c|c||c|c||c|c|}
					\hline
					$H$       							& \multicolumn{2}{c|}{\textbf{0.1}}& \multicolumn{2}{c|}{\textbf{0.3}} & \multicolumn{2}{c|}{\textbf{0.5}}  & \multicolumn{2}{c|}{\textbf{0.7}} & \multicolumn{2}{c|}{\textbf{0.9}}\\ \hline
					\textbf{Mean/CV}      				& Mean     & CV           & Mean        & CV       & Mean      & CV       & Mean      & CV & Mean      & CV \\ \hline
					$\sigma(Y_t) = \sqrt{Y_t + 0.1}$ 	& 0.774185342 &	0.062159457 &	0.782211975 &	0.015363114 &	0.775305642 &	0.053605636 &	0.765667823 &	0.022561751 &	0.776062568 &	0.061121985 \\ \hline
					$\sigma(Y_t) = Y_t + 0.1$			& 0.932824188 &	0.023154477	& 0.959352477 &	0.019764205 &	0.946670803 &	0.008803027 &	0.952432308 &	0.016014640 &	0.948353316&	0.008871172
					\\ \hline
					$\sigma(Y_t) = \sqrt{Y_t^2 + 1}$ 	& 0.707885444 &	0.093317545 &	0.715438258 &	0.077237936	& 0.695277007	& 0.053520175 &	0.720631067 &	0.041407711 &	0.729078909 &	0.085659766
					\\ \hline
				\end{tabular}%
			}
		\end{table}
	
		
		\begin{table}[]
			\centering
			\caption{Option prices using (\ref{Eq5-4})}
			\label{Tab5-2}
            \vspace{-3mm}
			\resizebox{\columnwidth}{!}{%
				\begin{tabular}{|c|c|c||c|c||c|c||c|c||c|c|}
					\hline
					$H$       							& \multicolumn{2}{c|}{\textbf{0.1}}& \multicolumn{2}{c|}{\textbf{0.3}} & \multicolumn{2}{c|}{\textbf{0.5}}  & \multicolumn{2}{c|}{\textbf{0.7}} & \multicolumn{2}{c|}{\textbf{0.9}}\\ \hline
					\textbf{Mean/CV}      				& Mean     & CV           & Mean        & CV       & Mean      & CV       & Mean      & CV & Mean      & CV \\ \hline
					$\sigma(Y_t) = \sqrt{Y_t + 0.1}$ 	& 0.79340973 & 0.07560649 &	0.81121348 &0.04028921 &0.78827183 &0.11421244 &0.76642501 &0.08935762& 0.7704734 & 0.13411309 \\ \hline
					$\sigma(Y_t) = Y_t + 0.1$			& 0.99910672 &0.09628926 & 0.95410606 & 0.16524115 & 0.97622451 & 0.06896021&  0.97074148& 0.10076119 & 1.013755924 & 0.10492516 \\ \hline
					$\sigma(Y_t) = \sqrt{Y_t^2 + 1}$ 	& 0.67871381 & 0.08759139& 0.69286223 & 0.09071164 & 0.66834204 & 0.10850252 & 0.69416225 & 0.09554705 & 0.707316469 & 0.07008638 \\ \hline
				\end{tabular}%
			}
		\end{table}
		

\newpage
\noindent
\textbf{Pricing options with volatility taking the form of  \textit{fCIR} process with time varying parameters} \\

\noindent
In this section, we perform some simulations of option prices under the fractional Heston model with time varying parameters. For this, the drift function is given by $f(t,z) = (\mu_t - \theta_t z^2)$, where $\theta_t = \theta>0$ and $\mu_t = c+\frac{\nu^2}{2\theta}\Big(1 - e^{-2\theta t}\Big)$. It follows that
$$
f(t,z) = \frac{\nu^2}{2\theta}\Big(1 - e^{-2\theta t}\Big) + (c-\theta z^2).$$

\noindent
We shall use $Z_0 = 1, ~\nu = 0.4, ~c=0.02, ~\theta = 1$. To keep positiveness of the stochastic process $(Z_t)_{t\geq 0}$ for all $H\in (0,1)$, we shall rather use its approximated stochastic process $(Z_t^\epsilon)_{t,\,\epsilon\geq 0}$ defined by (\ref{Eq2-1}), that is

$$
dZ^\epsilon_t = \frac{1}{2}f(t,Z^\epsilon_t)\Lambda_\epsilon(Z^\epsilon_t)dt + \frac{\sigma}{2}dW_t^H, ~~~~ Z_0^\epsilon = Z_0 > 0,
$$

\noindent
where the function  $\Lambda_\epsilon(z)$ is defined by
$$
\Lambda_\epsilon(z) = (z \mathbf{1}_{\{ z >0 \}} + \epsilon)^{-1}
$$

\noindent
with $\epsilon = 0.01$ for $H\leq 1/2$ and $\epsilon = 0$ for $H > 1/2$. As previously, the \textit{fBm} is simulated by using the formula (\ref{Eq5-10}) and (\ref{Eq5-11}). We perform again 100 trials for 500 simulations and 500 time-steps on the time interval $[0,1]$. We get the mean of option prices with their corresponding coefficient of variations for different volatility functions $\sigma(y)$ under the European-Binary option as given in Table \ref{Tab5-4} for direct estimations and in Table \ref{Tab5-5} by using (\ref{Eq5-4}).

		\begin{table}[]
			\centering
			\caption{Option prices using Direct Estimations}
			\label{Tab5-4}
			\resizebox{\columnwidth}{!}{%
				\begin{tabular}{|c|c|c||c|c||c|c||c|c||c|c|}
					\hline
					$H$       							& \multicolumn{2}{c|}{\textbf{0.1}}& \multicolumn{2}{c|}{\textbf{0.3}} & \multicolumn{2}{c|}{\textbf{0.5}}  & \multicolumn{2}{c|}{\textbf{0.7}} & \multicolumn{2}{c|}{\textbf{0.9}}\\ \hline
					\textbf{Mean/CV}      				& Mean     & CV           & Mean        & CV       & Mean      & CV       & Mean      & CV & Mean      & CV \\ \hline
					$\sigma(Y_t) = \sqrt{Y_t + 0.1}$ 	& 0.757738549 &	0.048177774 &	0.769114549 &	0.057692257 &	0.756162793 &	0.045562288 &	0.756665572 &	0.051234111 &	0.763148888&0.043265712 \\ \hline
					$\sigma(Y_t) = Y_t + 0.1$			& 0.932035897 &	0.012595508 &	0.934337494 &	0.022642941 &	0.933212125 &	0.024487 &	0.928706032	& 0.014969569 &	0.929103212 &	0.01457107   \\ \hline
					$\sigma(Y_t) = \sqrt{Y_t^2 + 1}$ 	& 0.770104152 &	0.088196662 &	0.782432528	& 0.062946479 &	0.75433847 & 0.069371091 &	0.746931996 &	0.072156192 &	0.75975843 &	0.084981952 \\ \hline
				\end{tabular}%
			}
		\end{table}
		
		
		\begin{table}[]
			\centering
			\caption{Option prices using (\ref{Eq5-4})}
			\label{Tab5-5}
			\resizebox{\columnwidth}{!}{%
				\begin{tabular}{|c|c|c||c|c||c|c||c|c||c|c|}
					\hline
					$H$       							& \multicolumn{2}{c|}{\textbf{0.1}}& \multicolumn{2}{c|}{\textbf{0.3}} & \multicolumn{2}{c|}{\textbf{0.5}}  & \multicolumn{2}{c|}{\textbf{0.7}} & \multicolumn{2}{c|}{\textbf{0.9}}\\ \hline
					\textbf{Mean/CV}      				& Mean     & CV           & Mean        & CV       & Mean      & CV       & Mean      & CV & Mean      & CV \\ \hline
					$\sigma(Y_t) = \sqrt{Y_t + 0.1}$ 	& 0.769174923 &	0.159481951 &	0.79459017 &	0.136648616 &	0.781942914 &	0.157116756	 & 0.747618003 &	0.12525256 &	0.755713234 &	0.06592363 \\ \hline
					$\sigma(Y_t) = Y_t + 0.1$			&  0.94650013 &	0.102404072 &	1.02769617 &	0.128530355	& 0.919334248 &	0.111971197 &	0.983793301&	0.095406694 &	0.88152163 &	0.101523439 \\ \hline
					$\sigma(Y_t) = \sqrt{Y_t^2 + 1}$ 	& 0.803170587 &	0.273211512 &	0.793796973 &	0.205160841 &	0.756164588 &	0.210899491 &	0.742696383	&0.203031148 &	0.759959966 &	0.198280955  \\ \hline
				\end{tabular}%
			}
		\end{table}

\section{Conclusion}

\noindent
In this paper, we have constructed an arbitrage-free and incomplete financial market model that consists of a risk-free asset with prices $A_t$ that verifies $dA_t = rA_tdt$ and the risky asset with price given as a geometric Brownian motion $dS_t = \eta S_t dt + \sigma(Y_t) S_t dB_t$. The volatility of infinitesimal return $dS_t/S_t$ given by $\sigma(Y_t)$ is a function of the generalised \textit{fCIR} process $(Y_t)_{t\geq 0}$ defined by $Y^2_t = Z^2_t \mathbf{1}_{[0, \tau)}$ with $ dZ_t = \frac{1}{2}f(t,Z_t)Z_t^{-1}dt + \frac{1}{2}\sigma dW_t^H,\,\, Z_0 >0, $ where $f(t,x)$ is a continuous function on $\mathds{R}_+^2$ that satisfies two mild conditions. We have investigated  different properties of each components of this financial market model.

\noindent
To support the results, we perform  simulations firstly when the volatility takes the form of Ornstein-Uhlenbeck process and recover the results in \citet{bezborodov2019option} and secondly, we consider the fractional Cox-Ingersoll-Ross process with time-varying parameters. We  have observed that option prices are more stable. Calibration of parameters will be subject to further investigations.


\bibliography {fHm_ArXiV}

\bibliographystyle {plainnat}

\end{document}